# AI Governance in the GCC States: A Comparative Analysis of National AI Strategies

**Mohammad Rashed Albous**       M.ALBOUS@KTECH.EDU.KW
*School of Business Management, Kuwait Technical College*
*Abu Halifa, Kuwait*

**Odeh Rashed Al-Jayyousi**       ODEHAJ@AGU.EDU.BH
*Department of Business Administration*
*Arabian Gulf University, Manama, Bahrain*

**Melodena Stephens**       MELODENA.STEPHENSB@MBRSG.AC.AE
*Academic Affairs, Mohammed Bin Rashid School of Gov't*
*Dubai, United Arab Emirates*

## ABSTRACT

Gulf Cooperation Council (GCC) states increasingly adopt Artificial Intelligence (AI) to drive economic diversification and enhance services. This paper investigates the evolving AI governance landscape across the six GCC nations, the United Arab Emirates, Saudi Arabia, Qatar, Oman, Bahrain, and Kuwait, through an in-depth document analysis of six National AI Strategies (NASs) and related policies published between 2018 and 2024. Drawing on the Multiple Streams Framework (MSF) and Multi-stakeholder Governance theory, the findings highlight a "soft regulation" approach that emphasizes national strategies and ethical principles rather than binding regulations. While this approach fosters rapid innovation, it also raises concerns regarding the enforceability of ethical standards, potential ethicswashing, and alignment with global frameworks, particularly the EU AI Act. Common challenges include data limitations, talent shortages, and reconciling AI applications with cultural values. Despite these hurdles, GCC governments aspire to leverage AI for robust economic growth, better public services, and regional leadership in responsible AI. The analysis suggests that strengthening legal mechanisms, enhancing stakeholder engagement, and aligning policies with local contexts and international norms will be essential for harnessing AI's transformative potential in the GCC.

## 1. Introduction

The Gulf Cooperation Council (GCC) states are leveraging Artificial Intelligence (AI) for economic diversification and societal progress. This research focuses on the six GCC member states: the United Arab Emirates (UAE), the Kingdom of Saudi Arabia (KSA), Qatar, Oman, Bahrain, and Kuwait, a burgeoning AI landscape with ambitious national strategies. This presents a unique opportunity to examine AI governance in a region navigating technological innovation within a distinct cultural context (Hendawy & Kumar, 2024; Rahayu et al., 2023). Some countries in the GCC have been early pioneers of the e-government movement, such as the UAE (Al Ali et al., 2023; Stephens, 2024), suggesting that they were poised for early adoption of AI. The GCC's financial resources and proactive digitalization position it at the forefront of AI adoption (Aqeel & Naser, 2024), envisioning AI as a catalyst for economic growth, enhanced public services, and establishing the GCC as a global leader in responsible AI (Al Bakri & Kisswani, 2024; Khan et al., 2022).

While AI governance is well-studied in the West, GCC-specific policymaking, implementation, and adaptation remain underexamined. Scholars have largely focused on pluralistic and democratized settings, leaving top-down governance structures underexplored. This lacuna is especially pressing given the GCC's rapid digitalization, substantial financial investments in AI, and distinct sociopolitical environments. By focusing on GCC states, this study extends current theoretical debates on AI governance and asks whether established policy frameworks hold or require adaptation in regions characterized by centralized authority and resource-driven innovation. Building on these insights, literature on AI governance in the GCC remains limited, with comparatively little empirical research on policy implementation, comparative analyses across member states, and the long-term social, economic, and political implications of AI.

To address these gaps, this research examines the following question: How do the GCC states govern AI, considering their challenges, shared aspirations, and alignment with international standards, particularly those of the EU? Section 2 provides an overview of AI governance, including research in the GCC (Section 2.1) and the identification of key research gaps, then explores critical dimensions of AI governance (Section 2.2), such as ethical considerations (Section 2.2.1), regulatory frameworks (Section 2.2.2), organizational aspects (Section 2.2.3), and AI governance capacities (Section 2.2.4). Section 3 outlines the methodology and presents the dual analytical frameworks (Section 3.3). Findings are reported in Section 4, followed by a discussion in Section 5 and concluding remarks in Section 6.

## 2. AI Governance: From Principles to Practice

This section provides an overview of the global landscape of AI governance, examines the current state of AI governance research in the GCC, and identifies key dimensions that shape AI policymaking. By detailing the historical and emerging approaches to AI governance worldwide, as well as the specific context of the GCC, this review lays the foundation for the empirical investigation that follows in subsequent sections.

### 2.1 The Global Landscape of AI Governance

Globally, AI governance is rapidly evolving. As of April 2024, over 94 countries have created or are implementing National AI Strategies (NASs), demonstrating widespread recognition of AI's transformative potential (Maslej et al., 2024). Figure 1 shows the global distribution of NASs: blue represents countries with a published strategy, pink those in development, and grey those without a published plan. While the surge in policy action is notable, national approaches differ significantly, reflecting each country's political structure, technological capacity, and strategic priorities.

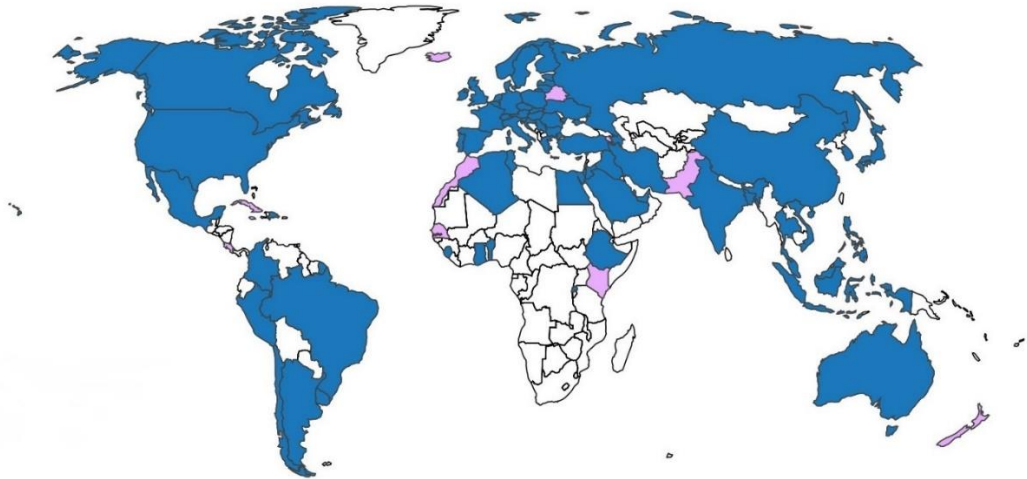

Figure 1: Countries with NASs (Maslej et al., 2024)

The OECD's Principles on AI (OECD, 2019; 2024) highlight human-centered values, accountability, transparency, and safety. However, implementation varies (Schmitt, 2022; Taeihagh, 2021). In May 2024, the European Union (EU) introduced its AI Act, which employs a risk-based regulatory framework to balance AI's potential benefits with possible harms (Madiega, 2021; World Economic Forum, 2024). Grounded in the European Charter of Fundamental Rights (2000), this legislation underscores the EU's commitment to safeguarding fundamental rights in AI governance. Further global developments include UNESCO's Recommendation on the Ethics of AI (UNESCO, 2021), the UN General Assembly's ethical AI principles (UN News, 2024), and the Council of Europe's efforts toward a legally binding AI treaty (Council of Europe, 2024).

Table 1 illustrates that while many nations, including GCC states, participate in non-binding ethical guidelines (e.g., UNESCO's framework), they frequently display caution about more stringent regulations. This fragmented global context has led to concerns over "ethicswashing", where lofty ethical declarations may lack robust enforcement (Schultz et al., 2024). At the same time, it offers countries flexibility to tailor governance to local conditions, a trend particularly relevant to resource-rich, rapidly modernizing regions like the GCC.

Table 1: GCC Countries Participation in International AI Governance Agreements

| GCC Countries | Non-Binding Agreements | | | Binding Agreements |
|---|---|---|---|---|
| | UNESCO Recommendations on the Ethics of AI (2021) | OECD AI Principles (adopted in 2019, amended on 2024) | UNGA (2024) | Council of Europe AI Treaty (2024) |
| KSA | ✓ | ✗ | ✓ | ✗ |
| UAE | ✓ | ✗ | ✓ | ✗ |
| Qatar | ✓ | ✗ | ✓ | ✗ |
| Oman | ✓ | ✗ | ✓ | ✗ |
| Bahrain | ✓ | ✗ | ✓ | ✗ |
| Kuwait | ✓ | ✗ | ✓ | ✗ |

Note * ✓ indicates participation in the agreement; ✗ indicates non-participation.

Several countries have prioritized NASs, with Canada leading the way in 2017, and the UAE following suit. However, there's a lack of global consistency in these approaches. For example, the United States has issued various reports and guidelines (National Science and Technology Council, 2023), but these federal initiatives haven't translated into cohesive state-level regulations. Similarly, China has its own ethical guidelines, stressing the importance of controllable, reliable AI that benefits humanity (State Council of the People's Republic of China, 2017).

## 2.2 AI Governance Research in the GCC: A Critical Review

In the GCC, each member state, Saudi Arabia, the UAE, Qatar, Oman, Bahrain, and Kuwait, has launched or updated an NAS. These strategies align AI development with national visions for economic diversification and global competitiveness. Figure 2 presents a timeline of NAS releases in the GCC from 2018 to 2024, highlighting how the region has swiftly incorporated AI into broader digital transformation agendas.

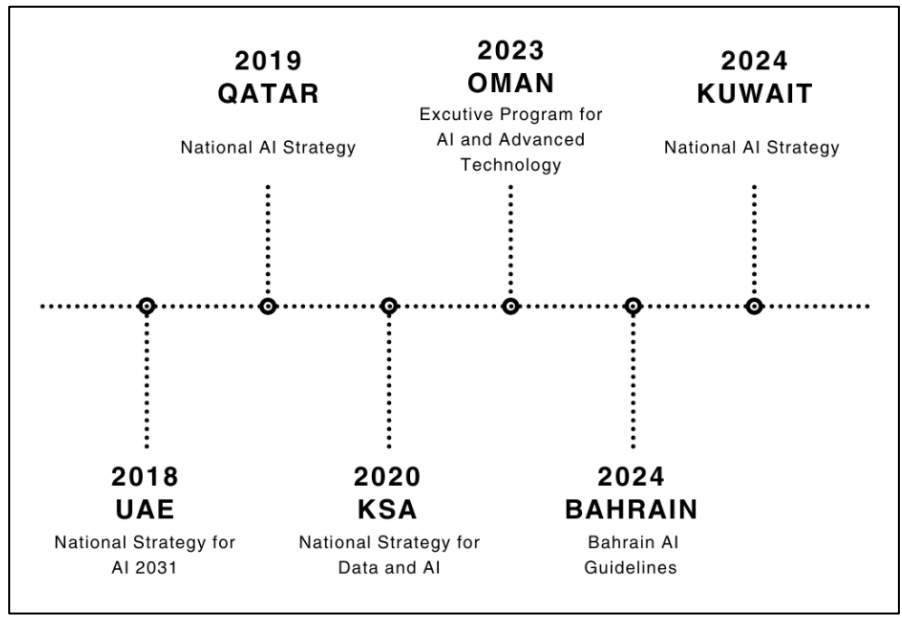

Figure 2: Timeline of NAS Adoption in the GCC

Despite comprehensive NASs, few research papers specifically address the intricacies of AI governance within the GCC. Existing research (Al-Barakati, 2021; Ashehri, 2019; Hendawy & Kumar, 2024) provides insights into the emerging AI governance landscape in the GCC and the wider Arab region, laying the groundwork for further investigation. Recent studies have also examined sector-specific applications of AI governance principles in the GCC, including the gas and oil industry (Al-Hajri et al., 2024), healthcare (Solaiman et al., 2024), and finance (Khan et al., 2022). This highlights the need for tailored governance approaches in each sector.

A recurring theme is the strong emphasis on leveraging AI for economic diversification and growth (Crupi & Schilirò, 2023), recognizing AI's potential in driving economic development and reducing reliance on traditional hydrocarbon industries. While ethical and social considerations are acknowledged, they may receive less attention compared to economic aspects, highlighting the need for further research into the ethical implications of AI in the region. These considerations should extend beyond general principles to include often-overlooked issues like the potential for political abuse of AI systems, the lack of diversity in the AI community, and the social and ecological costs associated with AI technologies (Hagendorff, 2020).

Some scholars have examined the role of cultural and religious factors in shaping AI governance in the GCC, with insightful analyses of how Islamic ethics can guide AI development and deployment (Rabbani et al., 2022; Rahayu et al., 2023). However, the broader context of AI ethics discussions and how this regional focus intersects with global trends should be acknowledged. There is a significant lack of representation from the Global South in these conversations (Roche et al., 2023). This raises concerns about ethnocentrism and the dominance of Global North perspectives, which may not adequately address the unique needs and challenges of regions like the GCC.

However, literature still lacks comparative, empirical, and implementation-focused studies on GCC AI governance. While existing work delves into sector-specific applications and general policy agendas, few studies thoroughly investigate how national strategies translate into regulatory practice, stakeholder engagement, or enforceable ethical standards. This gap is particularly salient given the GCC's rapid policy developments, strong government-led initiatives, and cultural-religious factors that may uniquely shape AI governance. Addressing these underexplored dimensions can clarify the interplay between policy intent and on-the-ground outcomes, thereby advancing a more nuanced understanding of AI governance across diverse GCC contexts.

## 2.3 Key Dimensions of AI Governance

Several broad dimensions surface consistently in AI governance scholarship and underpin both global and GCC-specific discourse (World Economic Forum, 2024). While these dimensions are typically addressed piecemeal, they collectively offer a framework for understanding how AI policy evolves in different contexts.

### 2.3.1 Ethical Considerations

Works such as Gabriel (2020) and Müller (2020) explore the philosophical foundations of AI ethics, while Sloane et al. (2022) delve into the practical challenges of implementing ethical principles in AI development. Corrêa et al. (2023) provide a comprehensive review of 200 AI guidelines worldwide, highlighting the global emphasis on ethical principles such as transparency, justice, non-maleficence, responsibility, and privacy. Different ethical approaches underpin these principles. For instance, principle-based frameworks emphasize fundamental ethical values, while rule-based frameworks focus on specific regulations (Manzoni et al., 2022; Sarker et al., 2024). Similarly, consequentialist ethics evaluates AI actions based on their outcomes, while deontological ethics emphasizes adherence to ethical rules and duties (Johnson et al., 2023; Zoshak & Dew, 2021). Analysing GCC AI policies through these lenses can reveal the underlying ethical assumptions shaping the region's approach to AI governance.

### 2.3.2 Regulatory Frameworks

Comprehensive legal and regulatory instruments are vital for governing AI development and deployment. The EU's AI Act (2024) has been particularly influential in this regard, introducing a risk-based approach that categorizes AI systems based on potential impacts and imposes proportionate requirements on higher-risk applications (Madiega, 2021; World Economic Forum, 2024). Importantly, the Act is not a standalone regulation but part of a broader legislative framework that includes the General Data Protection Regulation (GDPR), the Digital Services Act, and the Digital Markets Act, creating a comprehensive ecosystem for digital governance. This integrated model offers a valuable point of comparison for the GCC as it formulates AI governance strategies, highlighting the tension between "soft law" instruments and more binding regulations. While the GCC's focus on flexible, innovation-friendly guidelines can accelerate AI adoption, scholars caution that reliance on non-binding measures may risk inconsistent enforcement and "ethicswashing" (Schultz et al., 2024).

### 2.3.3 Organizational and Procedural Aspects

Effective AI governance requires not only ethical principles and regulations but also robust organizational and procedural mechanisms for practical implementation. These mechanisms include clearly defined roles and responsibilities, established review processes, and a culture of responsible AI development (Domingos, 2015). Examining how the GCC translates ethical aspirations and regulatory frameworks into concrete practices will provide insights into the effectiveness of AI governance in the region. AI governance typically involves various levels, including government agencies, private companies, and non-profit organizations (Outeda, 2024; Medaglia et al., 2024). A multi-stakeholder approach in decision-making is crucial, with representatives from government, industry, academia, and civil society (NIST, 2023).

### 2.3.4 AI Governance Capacities

Strong implementation capacities, skills, and data practices are vital for effective AI governance. AI capacity building demands a multifaceted approach, encompassing both technical skills development and data literacy (Agrawal et al., 2021). Investing in AI skills development, including education and training in areas such as machine learning, data science, and AI ethics, is essential (Acemoglu & Restrepo, 2022). Effective data management practices are also needed, including establishing clear data governance frameworks, implementing data quality controls, and ensuring data security and privacy (Janssen et al., 2023). Assessing GCC capacities clarifies readiness to harness AI's benefits and mitigate its risks.

## 3. Methodology

This research employs a document analysis approach to investigate the evolving landscape of AI governance in the GCC. Document analysis is defined as "a systematic and replicable technique for condensing large volumes of text into manageable categories using clear coding rules" (Neuendorf, 2017), making it particularly appropriate for this study for several reasons. First, it enables a systematic review of official government documents, reports, and policy papers, thereby offering empirical evidence on AI governance approaches in the GCC (Hendawy & Kumar, 2024). Second, it facilitates an encompassing view of multiple data sources, NASs, academic publications, media reports, and thus captures diverse perspectives. Third, examining documents published over a defined period (2018–2024) allows us to trace policy discourses and identify temporal trends in AI governance. Finally, the method proves especially suitable for the GCC context, where publicly available textual data may be the most substantial evidence of AI governance trajectories.

### 3.1 Document Selection

The document selection process prioritized a balance between breadth and relevance. The selection criteria focused on NASs from each GCC state, published between 2018 and 2024, explicitly addressing AI governance, and available in English or Arabic. Six NAS documents met these criteria. An additional 10 documents were examined but excluded due to lack of relevance. The primary analysis centered on these six NAS documents within the broader AI governance landscape of the GCC, along with a wider range of sources, such as privacy and cybersecurity laws, official policy statements, academic publications, and media reports. The search strategy encompassed both academic databases (OECD.ai, Scopus, Web of Science, and Google Scholar) and relevant government websites. Snowball sampling was also employed. Each document underwent a thorough quality assessment, considering factors such as the source's credibility, author credentials, and methodological rigor.

This study covers AI governance documents and policies up to January 2025. Any subsequent changes or newly issued frameworks fall outside our scope and may alter the governance landscape described here.

## 3.2 Analysis

Our analysis involved three key techniques:

**A. Content Analysis:** Systematic coding of the documents using a pre-defined codebook based on our research question and analytical frameworks, allowing us to identify key themes, concepts, and patterns in the data. The codebook is available in Table 12 at https://doi.org/10.5281/zenodo.14782333.

**B. Thematic Analysis:** Exploring the underlying meanings and relationships between the identified themes through iterative reading and interpretation of the data, guided by our research question and analytical frameworks.

**C. Term Frequency-Inverse Document Frequency (TF-IDF) Analysis:** Used to analyse the importance of words in our collection of NAS documents, determining the significance of a word to a document within a larger collection of documents (Manning, 2009). TF-IDF is calculated as: **TF-IDF = TF * IDF**. Where **TF** (Term Frequency) measures how frequently a term (t) appears in a document (d): **TF(t, d) = (Number of times term t appears in document d) / (Total number of terms in document d)**. And **IDF** (Inverse Document Frequency) measures the importance of a term across the document collection: **IDF(t) = (Total amount of documents / Number of documents with term t in it)**. Our analysis considered all instances where terms were discussed, regardless of whether they appeared in titles, informative statements, or normative statements (Lewis et al., 2020).

To prepare our NAS documents for analysis, we converted them into a text-based format using the pdfplumber Python library. For documents with information boxes displayed as images, we utilized a combination of Adobe's online OCR service and the Tesseract Python library. We estimate the error rate from the OCR process to be less than 5% and the OCR process was used for approximately 17% of the analysed corpora. To ensure quality and consistency across English and Arabic texts, we used professional translation services for all Arabic documents, followed by back-translation. We also developed a multilingual glossary of key AI terms in both English and Arabic.

## 3.3 Analytical Frameworks

To interpret the collected documents, we draw on two complementary frameworks, MSF and Multi-stakeholder Governance. These lenses help illuminate both the policy formation process and the enabling environment for stakeholder engagement. Their operationalization in this study is detailed below. Figure 3 illustrates how these frameworks intersect to yield a more holistic understanding of AI governance in the GCC.

### 3.3.1 Multiple Streams Framework (MSF)

MSF suggests policy change happens when problem, policy, and politics streams converge (Kingdon, 1984). While Western democracies see these streams shaped by pluralistic competition, in the GCC, top-down directives like Vision 2030 and UAE Centennial 2071 drive them differently. To understand this, the study analysed GCC AI documents, coding the problem stream by identifying challenges driving AI adoption (e.g., economic diversification), the policy stream by examining proposed solutions (e.g., new laws, ethics guidelines), and the politics stream by analysing references to political leadership and resource allocation that influence policy outcomes. This approach helps understand how AI governance emerges in the GCC's unique, resource-rich, non-pluralistic context.

### 3.3.2 Multi-stakeholder Governance

Multi-stakeholder Governance theory stresses the importance of diverse stakeholder participation, including government, industry, academia, and civil society, in shaping effective and ethical AI governance (Bevir, 2012; Roloff, 2008). In the GCC, a region characterized by strong central leadership and relatively new AI initiatives, such stakeholder involvement may be limited or predominantly government driven. Consequently, this research considers not only the existing level of stakeholder engagement but also the enabling environment that can foster broader participation.

A key component of this enabling environment is the legal and regulatory framework around data protection. Without robust Personal Data Protection Laws (PDPLs) to safeguard individuals' data rights, it is difficult to build the trust necessary for civil society, businesses, and individuals to meaningfully engage in AI governance. We therefore interpret PDPL robustness as an indicator of a government's commitment to multi-stakeholder collaboration. Such laws clearly signal a willingness to protect fundamental rights and establish transparent rules on data use, privacy, and accountability.

In assessing the presence and scope of PDPLs across GCC states, this analysis explores whether and how these laws address the unique challenges posed by AI technologies, including mechanisms for redress and enforcement. Additionally, we examine evidence of stakeholder engagement, such as private-sector partnerships, academic collaborations, and civil society, in NASs and policy documents. By evaluating both legal readiness (PDPLs) and practical engagement (stakeholder participation), we aim to determine the extent to which multi-stakeholder governance is actively fostered and the potential for inclusive, representative AI governance that accounts for diverse perspectives.

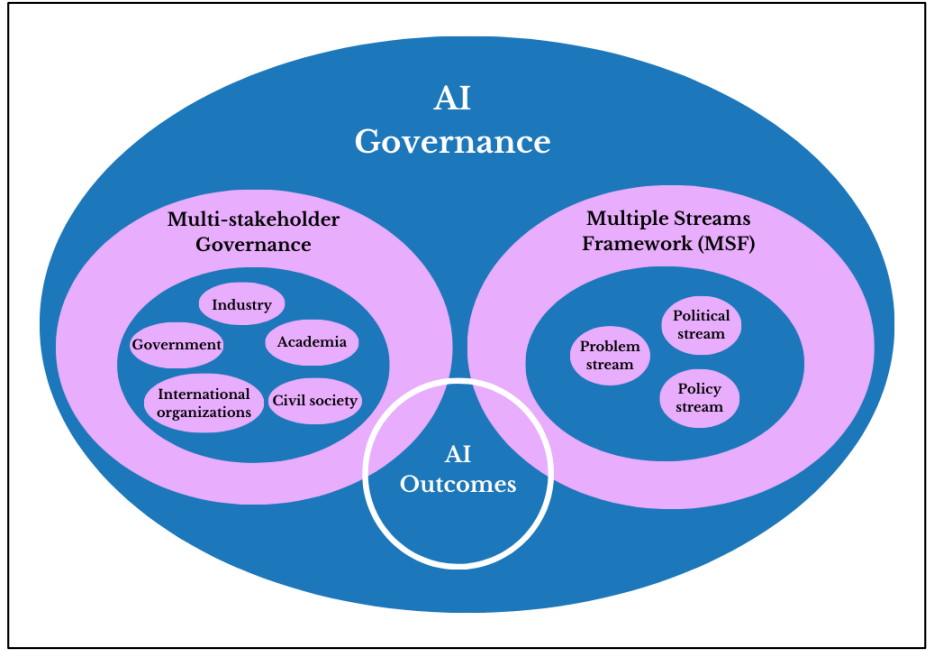

Figure 3: AI Outcomes: Multiple Streams and Stakeholder Dynamics

### 3.4 Limitations

First, the selection and interpretation of documents can be influenced by document-centric bias. Document-centric bias can miss on-the-ground realities and informal practices, limiting analysis depth. Although we mitigated this by involving multiple coders in the analysis, the subjective nature of qualitative research remains a factor. Two co-authors with extensive experience in qualitative research and expertise in AI governance and GCC policies served as inter-coders. Inter-coder reliability was assessed on the entire dataset using Cohen's kappa, yielding a value of 0.78, which indicates substantial agreement (Cohen, 1960). Any initial disagreements in coding were resolved through discussion and consensus among the three researchers.

Second, the reliance on textual data may not fully capture the nuances of AI governance practices as they are implemented. Additionally, while the frequency of terms in Table 4 and Figure 5 (using TF-IDF) may not be a perfect indicator, it could lead to overstating certain terms' importance while underrepresenting others that are contextually significant but mentioned less frequently.

Third, his study examines GCC AI frameworks' aims but not their long-term impacts. Longitudinal studies are needed to evaluate social, economic, and political impacts over time.

## 4. Findings

This section presents the research findings, exploring AI governance in the GCC states through the lens of two analytical frameworks: the MSF and the Multi-stakeholder Governance framework.

## 4.1 Application of Analytical Frameworks

This study employed two frameworks to systematically analyze the AI governance landscape in each GCC state: the MSF and the Multi-stakeholder Governance framework. At the same time, we recognize that research initiatives, awards, or the establishment of AI research centers alone do not inherently guarantee meaningful policy implementation; rather, they can be viewed as upstream enablers that shape practical outcomes. By situating these "soft" measures within the broader context of policy streams and stakeholder collaboration, the analytical frameworks help illuminate how AI governance policies in the GCC move from aspiration to execution.

### 4.1.1    Multiple Streams Framework (MSF)

The MSF analyzes how policy problems, solutions, and political factors converge to create windows of opportunity for policy change. The analysis reveals variations in how each GCC state approaches these streams.

- **UAE:** The UAE demonstrates a strong alignment of all three MSF streams. The problem stream is recognized as the need for economic diversification and enhanced public services. The policy stream is well-defined, with the National Strategy for Artificial Intelligence 2031, initiatives like the AI Council, and sector-specific initiatives like the UAE AI and Robotics Award. The establishment of MBZUAI, focused on talent development and research, supports these policies. The politics stream shows strong commitment, evidenced by the Office of Artificial Intelligence, Digital Economy and Remote Work Applications and significant investments in infrastructure and advanced LLM like Jais.
- **KSA:** Saudi Arabia exhibits a convergence of the problem and policy streams, with Vision 2030 as the key driver and the National Strategy for Data and Artificial Intelligence (and initiatives like the National AI Centre) articulating the policy stream. The politics stream shows commitment through investment in research and talent development and establishing SDAIA to lead AI initiatives. However, it may need further strengthening to fully integrate AI into the broader innovation ecosystem. This could involve fostering a more dynamic entrepreneurial environment and ensuring that stakeholder engagement extends beyond high-level partnerships to include a broader range of domestic actors.
- **Qatar:** Qatar showcases a connection between the problem and policy streams, driven by Qatar National Vision 2030 and with the policy stream embodied in their NAS and initiatives like the AI Committee. The successful integration of AI in digital government services, like Metrash, Hukoomi, and Baladiya/Oun, demonstrates effective policy implementation. However, the politics stream needs a stronger push to develop a comprehensive legal framework addressing ethical considerations, potentially through greater resource allocation and a more prominent political focus on AI governance.
- **Oman:** Oman's AI strategy, framed within Oman Vision 2040, demonstrates an alignment between the problem and policy streams. However, the politics stream appears weak, lacking the drive to translate strategy into action, requiring clear implementation plans and measurable goals.
- **Bahrain:** Bahrain demonstrates a nascent policy stream, with AI applied to enhance government services, such as AI-powered chatbots. However, the problem stream lacks clear articulation beyond improving efficiency. The absence of a comprehensive NAS suggests a weak politics stream, indicating a need for greater political attention and resource allocation to AI governance.
- **Kuwait:** Kuwait highlights a strong intention within the problem and policy streams, aligned with the "New Kuwait 2035" vision and a NAS that emphasizes human capital development. However, the politics stream lacks the necessary drive to translate the strategy into action, requiring a stronger political push, focusing on actionable steps, accountability, and measurable outcomes.

Overall, the MSF analysis reveals that GCC states vary in how strongly the three streams align. The UAE currently leads in weaving its AI strategies into broader political and economic frameworks, while the others display potential yet face differing gaps in political push, legislative mechanisms, or policy clarity. These variations directly inform how AI is governed and whether "windows of opportunity" in each country can produce concrete governance outcomes.

### 4.1.2   Multi-stakeholder Governance Framework

Analysis of AI governance frameworks across the GCC reveals a nascent but growing trend towards multi-stakeholder engagement. While government agencies currently play a dominant role, initiatives in UAE and KSA demonstrate an increasing effort to involve industry and academic stakeholders in policy discussions. However, the participation of civil society remains limited, suggesting a potential area for future development.

- **UAE:** The UAE stands out for its relatively robust multi-stakeholder approach, actively involving government bodies (like the Office of Artificial Intelligence, Digital Economy and Remote Work Applications and the AI Council), the private sector (with companies such as G42 and Hub71), and academic institutions (like MBZUAI). Initiatives like the UAE AI Ecosystem foster concrete collaboration. The UAE's PDPL (2021), with its emphasis on user control, data minimization, and transparency, is a crucial indicator of the government's commitment to fostering a multi-stakeholder ecosystem by establishing a foundation of trust and enabling responsible data sharing.
- **KSA:** Saudi Arabia, led by the SDAIA, demonstrates a commitment to multi-stakeholder partnerships, particularly on the international stage, as exemplified by the Global AI Summit. The KSA's PDPL (2023) is a significant step towards creating an enabling environment for multi-stakeholder collaboration by providing a framework for data protection. However, to further develop a robust multi-stakeholder ecosystem, KSA should focus on deepening engagement with domestic stakeholders, including startups, researchers, and civil society, and developing more robust accountability frameworks within its AI governance.
- **Qatar:** Qatar's NAS acknowledges the importance of multi-stakeholder collaboration yet lacks concrete implementation mechanisms. While its data protection law (2016) provides a basic foundation, it needs to be more comprehensively integrated into a broader AI governance framework to foster trust and encourage broader participation in the AI ecosystem.
- **Oman:** Oman recognizes the need for collaboration between government, academia, and industry in its NAS but lacks precisely defined roles and responsibilities. Oman's PDPL (2022) is a positive step, but it needs further integration with a comprehensive AI governance strategy and stronger accountability mechanisms to create a truly multi-stakeholder environment.
- **Bahrain:** Bahrain's AI governance appears to be primarily government-driven, with the Information and eGovernment Authority as a key player. Limited engagement with the private sector, academia, and civil society is evident. While its PDPL (2019) provides a starting point, it needs to be more thoroughly integrated with a comprehensive AI governance framework to build trust and encourage broader stakeholder participation.

- **Kuwait:** Kuwait acknowledges the importance of collaboration in its NAS. However, the strategy lacks specific details. The 2024 updated Data Privacy Protection Regulation, while a step forward, only applies to telecommunications and IT service providers and excludes the government's own data practices. This limited scope hinders the development of robust multi-stakeholder governance. To foster a truly collaborative environment, Kuwait needs to extend data privacy protections across all sectors, including the government, through comprehensive legislation.

From a multi-stakeholder standpoint, UAE and KSA are emerging leaders, actively involving industry and academia. Yet civil society engagement remains underdeveloped region-wide, reflecting the top-down policy context. This theoretical lens helps us see where stakeholder participation lags, particularly among community groups, non-profits, and smaller tech start-ups, and why bridging that gap is crucial for credible, inclusive AI governance in line with global standards.

## 4.2 Comparative Analysis of AI Governance in the GCC States

Our document analysis reveals a dynamic and evolving landscape of AI governance in the GCC, with common aspirations and shared challenges, yet unique nuances in each state's approach. We structure our key findings along the lines of our research question:

### 4.2.1   The Current State of AI Governance in the GCC

- **Ambitious and Robust National AI Strategies**

All GCC states have formulated comprehensive NASs outlining ambitious plans for AI development and adoption, emphasizing ethical and responsible AI use. Table 2 shows the focus of each strategy.

Table 2: GCC NASs Focus

| Country | GCC NASs Key Points/Themes |
|---------|----------------------------|
| KSA | Focus on data and AI, ambition to be a global leader, emphasis on skills development, investment, research, and creating a favourable ecosystem. |
| UAE | Aims to be a world leader in AI by 2031, focus on investment in AI, developing an AI ecosystem, attracting talent, and ensuring strong governance. |
| Qatar | Stresses the importance of AI for the future, focuses on developing AI capabilities in education, data access, employment, business, research, and ethics. |
| Oman | Vision to establish a unique position in AI and advanced technologies, focus on collaboration, AI adoption in key sectors, and human-centric approach to AI governance. |
| Bahrain | Focus on using AI to improve government services and business efficiency, emphasis on ethical AI use, and initiatives in AI training, research, and development. |
| Kuwait | Aspires to be a leader in AI innovation, plans to use AI to improve various sectors, and aims to create a robust AI ecosystem. |

- **Emerging Regulatory Frameworks**

AI governance in the GCC has gained significant momentum, particularly in 2024. The number of AI governance documents, such as NASs and ethical guidelines, has increased since 2018, jumping to four in 2024, marking a significant surge. As illustrated in Figure 4, this trend underscores how AI-related policy instruments in the region have steadily multiplied over the years, reflecting a pronounced focus on strategic, non-binding guidance. While AI provisions within GCC laws have seen more modest growth since their emergence in 2019, with only one new provision in 2024, the increase in governance documents indicates a focus on non-binding instruments.

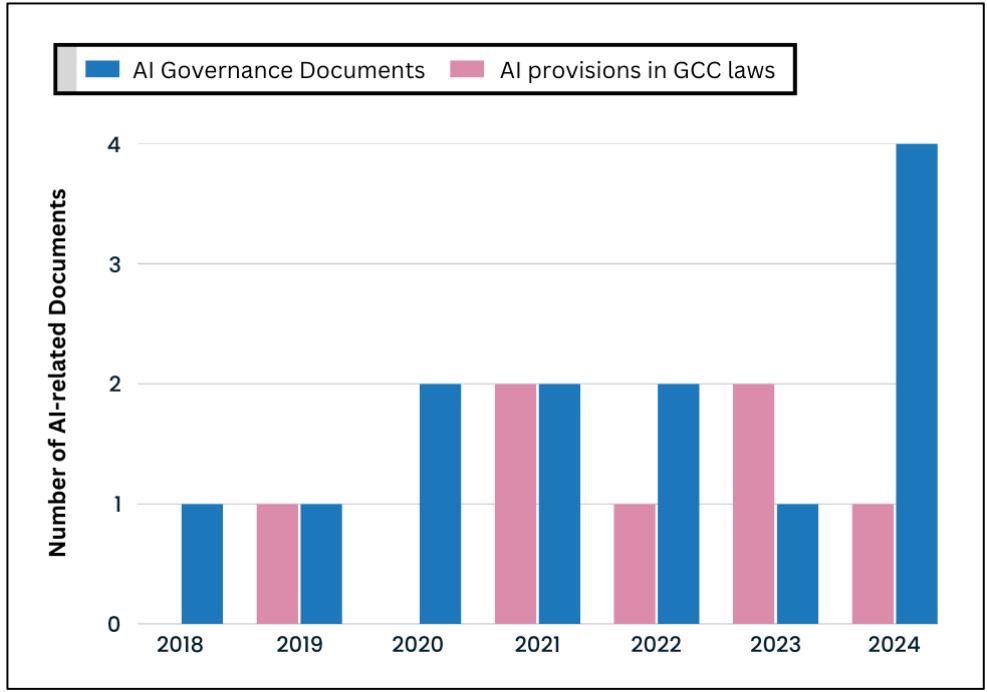

Figure 4: Growth in AI Related Documents in the GCC

While comprehensive, AI-specific legislation does not yet exist across the GCC, several states are proactively integrating AI-related provisions into their existing legal frameworks. These provisions primarily address concerns surrounding data privacy, algorithmic transparency, security, and potential liability arising from AI systems.

Notably, the Dubai International Financial Centre (DIFC) updated its Data Protection Regulations in September 2023, introducing the first enacted regulation in the MEASA region on processing personal data via autonomous and semi-autonomous systems such as AI or generative AI, and machine learning technology. This landmark development, particularly Regulation 10, signals a growing commitment to addressing the unique challenges posed by AI within existing legal structures. Table 3 offers a broader overview of such AI-related provisions within existing GCC laws.

Table 3: Overview of AI-related Provisions in Existing GCC Laws

| Country | Law/Regulation |
| --- | --- |
| UAE | UAE Federal Personal Data Protection Law (2021) |
| | Cybercrime Law (Federal Decree-Law No. 34 of 2021) |
| KSA | KSA Personal Data Protection Law (2023) |
| Oman | Royal Decree No. 6/2022 promulgating the Personal Data Protection Law (2022) |
| Qatar | Law No. 13 of 2016 on the Protection of Personal Data |
| | National Cyber Security Agency (NCSA) Guidelines for Secure Adopting and Usage of AI |
| Bahrain | Personal Data Protection Law (PDPL), Law No. 30 of 2019 |
| Kuwait | Data Privacy Protection Regulation No. 26 of 2024 |

- **Prioritizing Ethical and Responsible AI**

GCC NASs uniformly stress ethical considerations for socially beneficial AI. Table 4 shows each principle's document frequency (df), how many NASs (out of six) reference it, and a binary TF-IDF score. For clarity, the methodology behind these calculations (i.e., TF = 1 if a principle is mentioned, IDF = LN(6 / df), etc.) is detailed in Section 3.2 of this paper. By comparing the average TF-IDF values, we can distinguish between more broadly acknowledged principles (e.g., Privacy & Data Protection) and those that are comparatively rare (e.g., Benefit to Humanity). This approach underscores both the broad ethical consensus across the GCC and some unique priorities embraced by individual states.

Table 4: Frequency of Ethical Principles in GCC NASs Documents

| Ethical Principle | Doc. Freq. (df) | $IDF = LN(6 / df)$ | Countries Mentioned | Avg. TF-IDF |
| --- | --- | --- | --- | --- |
| Privacy & Data Protection | 6 | 0.00 | KSA, UAE, Kuwait, Bahrain, Qatar, Oman | 0.00 |
| Accountability | 5 | 0.18 | KSA, UAE, Kuwait, Bahrain, Qatar | 0.15 (approx.) |
| Transparency | 5 | 0.18 | KSA, UAE, Kuwait, Bahrain, Oman | 0.15 (approx.) |
| Fairness & Non-discrimination | 4 | 0.41 | KSA, UAE, Bahrain, Oman | 0.35 (approx.) |
| Human Oversight & Control | 3 | 0.69 | UAE, Bahrain, Oman | 0.58 (approx.) |
| Robustness & Safety | 3 | 0.69 | UAE, Kuwait, Oman | 0.58 (approx.) |
| Sustainability & Environmental Friendliness | 2 | 1.10 | UAE, Bahrain | 0.90 (approx.) |
| Human-cantered Values | 2 | 1.10 | UAE, Oman | 0.90 (approx.) |
| Collaboration & Inclusivity | 1 | 1.79 | KSA | 1.79 (approx.) |

| Ethical Principle | Doc. Freq. (df) | IDF = LN (6 / df) | Countries Mentioned | Avg. TF-IDF |
|---|---|---|---|---|
| Benefit to Humanity | 1 | 1.79 | Qatar | 1.79 (approx.) |

- **Human-Centric AI**

While the concept of AI as a tool for human empowerment is present in GCC NASs, its emphasis is less pronounced compared to other themes like economic growth and innovation. The documents often highlight the importance of using AI to enhance human capabilities and well-being, but the explicit prioritization of maintaining human agency and control in decision-making is less prominent. This suggests a potential area for further development in GCC AI governance, ensuring that AI technologies are harnessed to truly serve and benefit the population.

In this analysis, "human-centric themes" refer to instances in the GCC NAS documents where the primary focus is on ensuring that AI technologies serve human needs, enhance human capabilities, and promote overall human well-being. Table 5 illustrates the presence of human-centric themes across NASs, alongside other prominent priorities.

Table 5: Quote Examples from NAS`s on Human-Centric themes

| Country | Human-Centric Themes in the GCC NASs |
|---|---|
| UAE | "We aim to give this human potential the best opportunities to nourish and flourish…" |
| KSA | "Skills initiatives aim to enhance KSA's human capital…" |
| Oman | "Pillar 4: Governance of AI and Advanced Technologies for a Human-Cantered Vision…" |
| Qatar | "Qatar National Vision 2030 in itself organized around four pillars: economic, social, human, and environmental…" |
| Bahrain | "AI systems should not be used to replace human judgment and decision-making. Humans should always be in control of AI systems and their use" |
| Kuwait | "Invest in human capital development and promote lifelong learning…" |

## 4.2.2 Key Challenges in Governing AI

- **Data Challenges**

Although not always explicitly labeled as "data scarcity" or "data quality", many documents indirectly address challenges related to data access and suitability for AI development in the GCC. This suggests an underlying awareness of these issues, even if they are not always explicitly stated. Despite promising efforts like the UAE's development of Jais, a 13-billion parameter LLM trained on a vast Arabic and English dataset, and the potential of LLMs to generate synthetic data to improve model performance in data-scarce scenarios, data challenges persist. Table 6 provides quote examples illustrating these data challenges, which may include references to the need for more data sharing, concerns about data privacy and security, or the importance of developing local datasets tailored to the GCC context.

Table 6: Quote Examples from NAS Documents Illustrating Data Challenges

| Country | Data Challenges in GCC NASs |
|---------|------------------------------|
| UAE | "The UAE realizes that the oil of the future is data and will invest into creating a robust data infrastructure. The UAE's ambition is to create a data-sharing program, providing shared open and standardized AI-ready data, collected through a consistent data standard" |
| KSA | "One of the central issues where government action is expected will be around data sharing and open data strategies" |
| Qatar | "Data Access is Paramount: develop data governance rules and guidelines that facilitate broad access to and sharing of data…" |
| Oman | "Facilitating access to national data by creating a platform for national data management, publishing open datasets, and enhancing the integration of government data systems" |
| Bahrain | Bahrain's AI strategy doesn't explicitly address data scarcity or quality |
| Kuwait | "Develop policies and standards to govern data security, privacy, quality, and access. Create secure ways for government, researchers, and businesses to share data while protecting privacy. Implement central data storage, ensure data quality, and use strong security to protect sensitive data…" |

- **Skills Gap and Capability Building**

GCC NASs consistently acknowledge the pressing reality of a global shortage of AI talent and expertise. This skills gap poses a significant obstacle to the region's ambitions for AI-driven transformation. While the UAE has taken a commendable lead in addressing this challenge by establishing the MBZUAI in 2019, a pioneering institution dedicated to AI education and research, offering specialized master's and Ph.D. programs, the demand for AI skills continues to outpace supply globally. Our analysis of GCC NAS documents further underscores this challenge, revealing a consistent emphasis on the need for capability building.

While frequency alone may not be a perfect indicator of actual needs, it does provide insights into which capabilities are emphasized and prioritized within the policy discourse. This analysis assumes that the more frequently a capability is mentioned, the more likely it is to be perceived as crucial for AI development and adoption in the region. However, it's important to acknowledge that other factors, such as the specific context of each country and the stage of AI development, also influence capability needs. Figure 5 provides a visual representation of the specific AI capabilities deemed most crucial based on their frequency of mention in these strategic plans. By pinpointing these prioritized areas, policymakers can guide targeted investments in education and training, ensuring that the GCC's workforce is equipped to harness the full potential of AI.

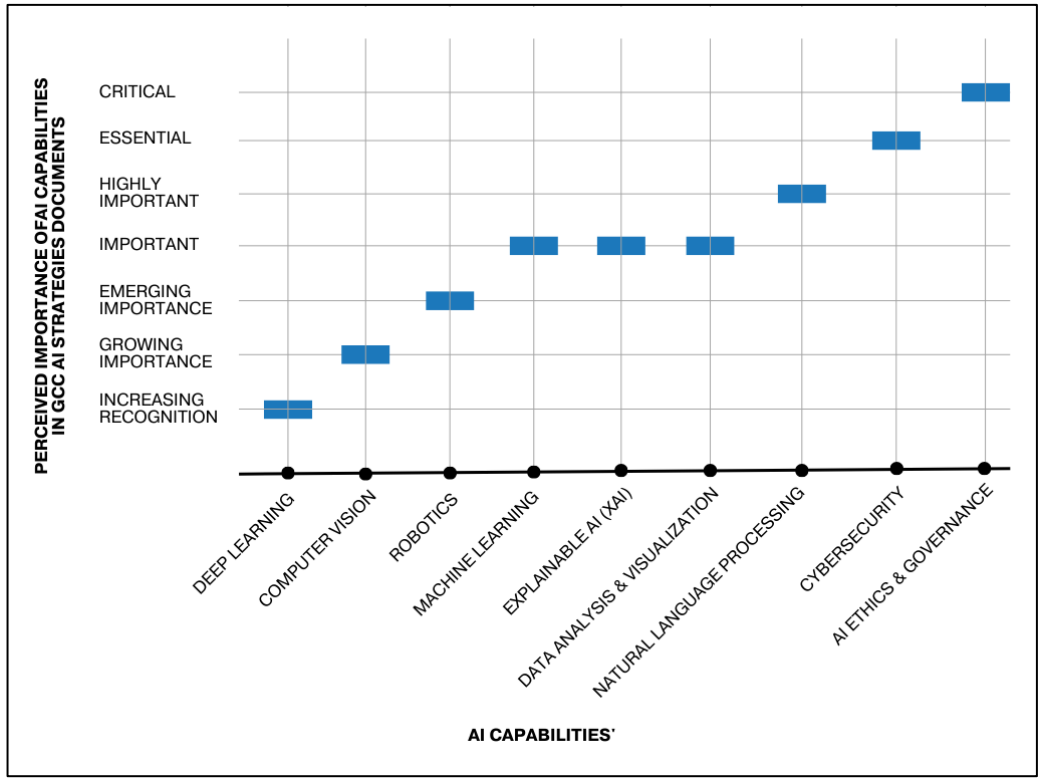

Figure 5: Bar Chart showing the Perceived Importance of different AI Capabilities in GCC NASs Documents

The bar chart in Figure 5 provides insights into the perceived importance of various AI capabilities within the context of GCC NASs documents. It's clear that AI Ethics & Governance is positioned as the most critical capability, underscoring the region's focus on responsible AI development and deployment. Cybersecurity is also deemed essential, reflecting the growing concern about safeguarding AI systems and data.

Natural Language Processing (NLP) is regarded as highly important, likely due to its potential applications in various sectors, including customer service and communication. Following closely, Data Analysis & Visualization and Explainable AI (XAI) are seen as important capabilities, suggesting a need for transparency and understanding in AI-driven decision-making. Machine Learning is also recognized as an important capability.

While Computer Vision is recognized as having growing importance, foundational capabilities such as Deep Learning appear to still be gaining traction, suggesting that the region is steadily building its AI expertise from the ground up.

- **Societal and Cultural Considerations**

At least one GCC country, Qatar, explicitly states that "the framework to be developed must be consistent with Qatari social, cultural, and religious norms. While this signals a growing awareness of the importance of cultural sensitivity in AI development and deployment across the region, it also highlights the inherent challenge of aligning rapidly advancing AI technologies with diverse and evolving cultural landscapes. Furthermore, the UAE's endeavour to develop and publish its LLM (Jais), mentioned earlier, demonstrates a commitment to addressing the unique cultural and linguistic needs of the region.

The imperative to align AI systems with the region's unique cultural values presents a significant challenge. Islamic principles, deeply embedded in GCC societies, offer a framework for ethical AI development. For instance, the emphasis on privacy (satr al-'awrat) can inform data governance, while social justice (adl) can guide the creation of fair and equitable AI systems. However, potential conflicts arise with technologies like facial recognition, which may clash with traditional values regarding modesty and personal space. Furthermore, the diverse interpretations of Islamic principles across the GCC necessitate a nuanced approach to AI governance, one that acknowledges and respects local contexts. Engaging with the literature on Islamic Ethics and AI can provide valuable guidance in navigating these complexities.

### 4.2.3  Shared Aspirations

- **Economic Diversification**

All GCC NASs emphasize the role of AI in diversifying their economies and reducing reliance on hydrocarbons. This is reflected in the Global Diversification Index Report (2024), which highlights their progress in this area, driven by post-pandemic reforms and incentives to invest in new technology sectors, particularly AI.

For example, the UAE aims for AI to generate up to $91 billion in extra growth (UAE National Strategy for Artificial Intelligence 2031, 2018). Similarly, Saudi Arabia has committed $20 billion to AI investment by 2030 through its Vision 2030 and National Strategy for Data and Artificial Intelligence (SADAIA, 2024), focusing on sectors like education, healthcare, energy, and transportation (NSDAI, 2020). Qatar has also allocated significant resources, with an incentive package worth $2.5 billion (Zawya, 2024).

These initiatives demonstrate the strategic importance of AI in the GCC's vision for a sustainable and prosperous future. While precise figures on AI investment for each GCC country are not always readily available, the trend clearly indicates a strong commitment to leveraging AI for economic transformation.

- **Enhanced Public Services**

The GCC states aspire to leverage AI to revolutionize public service delivery in sectors like healthcare, education, and transportation, aiming to streamline processes, personalize services, and improve efficiency and quality. Table 7 showcases concrete examples of AI applications envisioned in GCC documents, highlighting the region's commitment to leveraging AI for the public good.

Table 7: Examples of AI Applications in Public Services Envisioned in GCC NASs Documents

| Country | Applications of AI in GCC Public Services |
|---------|-------------------------------------------|
| UAE | "The UAE is… adjusting transport timetables to respond to incidents, using AI sensors for smart traffic, deploying facial recognition to monitor driver fatigue and introducing chatbots to improve customer service" |
| KSA | SDAIA created Estishraf to use government data to support decision-making, address key priorities and improve citizen's lives through better government services |

| Country | Applications of AI in GCC Public Services |
|---------|-------------------------------------------|
| Qatar | "Qatar already has a strong digital government, with services such as Metrash, Hukoomi, and Baladiya/Oun15. It is befitting to augment these services with AI capabilities…" |
| Oman | "Applying artificial intelligence in public services such as health, education and smart government applications that deal directly with citizens, residents and investors…" |
| Bahrain | "The Information and eGovernment Authority is working on the chatbot project" |
| Kuwait | "Implementing AI-powered automation in administrative processes, introducing AI chatbots for citizen support, and utilizing data analytics for decision making…" |

- **Regional Leadership in AI**

The GCC states aspire to be more than just adopters of AI; they aim to become global leaders in its ethical and responsible development and deployment. This ambition is evident not only in their NASs but also in various policy documents and official statements, reflecting a desire to actively shape the global AI discourse and set a benchmark for responsible innovation. This proactive stance underscores the GCC's recognition of AI's transformative potential and its determination to leverage this technology for regional advancement while upholding ethical considerations. Table 8 provides illustrative quotes from GCC AI documents supporting this observation.

Table 8: Illustrative Quotes from GCC NAS Documents on the GCC's Vision for AI Leadership

| Country | Quotes from GCC NAS Documents | Potential Leadership Areas |
|---------|-------------------------------|----------------------------|
| UAE | "The UAE has a vision to become one of the leading nations in AI by 2031" | AI talent development, AI ecosystem building |
| KSA | "KSA aim becoming one of the leading economies utilizing and exporting Data & AI after 2030" | AI research and development, data governance |
| Qatar | "Qatar is poised to play a leadership role in an AI+X Future" | AI in education and public services |
| Bahrain | "The vision of Bahrain's leadership to employ modern technologies such as Artificial Intelligence (AI) has improved government services and contributed to Bahrain's digital achievements" | AI in government services, practical AI applications |
| Kuwait | "The Kuwaiti Government is committed to leading by example in the responsible adoption and governance of AI" | Human capital development in AI |

While most GCC states aspire to leadership in AI, their strengths lie in different areas. The UAE, with initiatives like the Office of Artificial Intelligence, Digital Economy and Remote Work Applications, AI Council, the UAE AI Ecosystem, and MBZUAI, is a frontrunner in AI talent development and ecosystem building. Saudi Arabia, with its significant investments, is poised to lead in AI research and development. Instead of competing, the GCC should embrace a collaborative approach, leveraging these individual strengths. Establishing a regional AI hub would facilitate knowledge sharing, joint research, and shared infrastructure development. This synergistic strategy, where each nation contributes its unique capabilities, will prevent fragmented efforts and resource duplication. By working together, the GCC can accelerate innovation and amplify its collective strengths, positioning the region as a whole as a unified and powerful global AI leader.

### 4.2.4    Comparison to International Standards

While the EU AI Act employs detailed risk tiers (Articles 6, 14, etc.) for high-risk systems, GCC states predominantly lean on non-binding, innovation-focused guidelines, resulting in varying degrees of data protection and enforcement rigor. Indeed, each GCC country has enacted personal data protection laws, but the extent to which they match the EU's GDPR-level stringency varies, with the UAE's DIFC regulations coming closest. These parallels and divergences nonetheless underscore a readiness among GCC governments to engage with global frameworks while tailoring AI governance to local contexts, consistent with prior studies indicating the region's preference for flexible, innovation-friendly approaches.

- **Alignment with EU Principles**

The emerging AI governance frameworks in the GCC countries show a clear alignment with the key ethical principles outlined in the EU's AI Act. This reflects the growing influence of global standards on AI development. This alignment not only highlights the GCC's dedication to responsible AI development but also creates opportunities for collaboration and knowledge sharing with the EU in this vital field. A comprehensive comparative analysis of these shared ethical principles across GCC NAS documents and the EU AI Act is presented in Table 9.

Table 9: Similarities in Ethical Principles between GCC NAS Documents and the EU AI Act

| Ethical principles | GCC NAS Documents | EU AI Act | Shared Emphasis |
|---|---|---|---|
| **Human Agency and Oversight** | Emphasizes human responsibility and control in AI systems | Requires human oversight for high-risk AI systems (Article 14) | Both stress the importance of human control and responsibility in AI to ensure alignment with human values and prevent harm. The GCC NAS documents also emphasize preserving human autonomy in AI interactions |
| **Fairness and Non-discrimination** | Promotes equitable treatment and avoidance of bias in AI systems | Prohibits AI systems that discriminate based on protected characteristics (Article 10) | Both aim to prevent bias and ensure fairness and equality in AI applications |
| **Transparency and Explainability** | Encourages transparency in AI systems and the ability to explain their decision-making processes | Requires transparency and explainability for high-risk AI systems (Article 13) | Both recognize the need to understand how AI systems work and the reasons behind their decisions |
| **Privacy and Data Protection** | Emphasizes the protection of personal data and | Requires AI systems to comply with data protection laws like GDPR (Article 9) | Both prioritize protecting individuals' privacy and data in AI development and use |

| Ethical principles | GCC NAS Documents | EU AI Act | Shared Emphasis |
|---|---|---|---|
| | privacy in AI systems | | |
| Safety and Security | Promotes the development of safe, secure, and reliable AI systems | Requires high-risk AI systems to meet safety and security standards (Article 15) | Both emphasize building AI systems that are safe, secure, and operate reliably |
| Accountability | Encourages accountability for the development and deployment of AI systems | Establishes clear lines of accountability for providers and users of AI systems, especially for high-risk applications (Articles 6 & 29) | Both recognize the importance of holding individuals and organizations responsible for the impacts of AI systems |
| Sustainability | Explicitly promotes environmentally responsible AI development and use | Indirectly addressed through focus on human oversight and fundamental rights | Both frameworks, though with different focuses, contribute to sustainable AI development |

- **Differences with EU Principles**

While there's a notable convergence on core ethical principles, the GCC's approach to AI regulation is currently less prescriptive and more focused on promoting innovation compared to the EU's risk-based approach. The EU AI Act, with its tiered regulatory framework, prioritizes preemptive measures to mitigate potential risks associated with AI systems. In contrast, the GCC, as reflected in various policy documents and NASs, emphasizes flexibility and adaptability to foster a conducive environment for AI development and adoption. Table 10 provides a detailed comparative analysis of these differences in ethical principles across GCC NAS documents and the EU AI Act. This divergence underscores the GCC's unique approach to balancing innovation and risk management in the context of AI governance.

Table 10: Differences in Ethical Principles between GCC NAS Documents and the EU AI Act

| Ethical principles | GCC NAS Documents | EU AI Act | Key Difference |
|---|---|---|---|
| Scope of Human Agency and Oversight | Promotes a general principle of human agency and oversight for all AI systems | Primarily focuses on human oversight for high-risk AI systems (Article 14) | GCC NAS Documents emphasize human involvement in all AI, while the EU AI Act prioritizes high-risk applications |
| Specificity of Fairness and Non-discrimination | Provides general guidance on fairness and avoiding bias | Sets specific legal requirements for avoiding discrimination, particularly regarding protected characteristics (Article 10) | GCC NAS Documents offer broader ethical guidance, while the EU AI Act provides concrete legal obligations |

| Ethical principles | GCC NAS Documents | EU AI Act | Key Difference |
|---|---|---|---|
| **Application of Transparency and Explainability** | Encourages transparency as a general principle for all AI systems | Mandates transparency and explainability for high-risk AI systems (Article 13) | GCC NAS Documents promote transparency broadly, while the EU AI Act requires it specifically for high-risk systems |
| **Depth of Privacy and Data Protection** | Provides general guidelines on data protection in AI | Sets specific legal requirements for data processing, aligned with GDPR (Article 9) | The GCC NAS Documents offer general guidance, while the EU AI Act enforces specific data protection laws. However, the UAE and KSA have established their own personal data protection laws, similar to the GDPR |
| **Enforcement of Safety and Security** | Promotes safety and security as general principles for AI development | Sets specific safety and security standards for high-risk AI systems (Article 15) | GCC NAS Documents encourage safe development, while the EU AI Act mandates compliance for high-risk applications |
| **Specificity of Accountability** | Provides general guidance on accountability in AI development and deployment | Establishes specific legal requirements for identifying and addressing responsibility, particularly for high-risk systems (Articles 6 & 29) | GCC NAS Documents offer broader ethical considerations, while the EU AI Act defines legal liabilities |
| **Explicitness of Sustainability** | Explicitly mentions sustainability as a key principle for environmentally responsible AI | Addresses sustainability indirectly through other principles like human oversight and fundamental rights | GCC NAS Documents directly address environmental concerns, while the EU AI Act incorporates them indirectly |

- **Opportunities for Collaboration & Harmonization**

The analyzed documents reveal potential areas for fruitful collaboration between the GCC and the EU on AI governance. For instance, several GCC NASs explicitly express the desire to partner with international organizations and leading AI hubs to exchange knowledge and expertise. However, while progress is evident, there remains room for further harmonization of AI governance frameworks both within the GCC itself and with international standards. Such harmonization could facilitate cross-border AI collaboration, attract investment, and enhance the region's global standing in the AI arena. Table 11 illustrates these potential collaboration opportunities.

Table 11: Potential Synergies in AI Development

| Opportunities for Collaboration | Collaboration Aspects |
|---|---|
| AI Talent Development and Research | Both the EU and GCC aim to address AI skill shortages by expanding educational programs, research partnerships, and faculty/student exchanges. The UAE's MBZUAI and KSA's SDAIA-led initiatives align with Europe's established AI research networks, enabling cross-border collaborations that foster expertise in culturally adaptive AI. Joint scholarships, summer schools, and dual-degree programs could strengthen human capital pipelines on both sides. |
| Regulatory Frameworks and Ethical AI | The EU's AI Act provides a risk-based template that can inform GCC AI regulations. By sharing best practices on compliance, enforcement mechanisms, and culturally responsive ethical guidelines, the EU and GCC can co-develop frameworks that protect rights while encouraging innovation. Saudi Arabia's NSDAI and the UAE's ethical governance initiatives, for instance, can benefit from Europe's GDPR experience, ensuring robust yet flexible legal structures. |
| Data Sharing and Standards Development | Both the EU AI Act and GCC NAS documents emphasize data governance, quality, and security. Collaborative efforts could focus on developing open-data standards, privacy safeguards, and AI interoperability protocols. Projects like the UAE's ambition for AI-ready data infrastructure or Qatar's commitment to broad data access intersect well with Europe's ethos on data sharing and the EU's work on standardizing cross-border digital flows. Jointly adopted frameworks would streamline global AI projects and foster mutual trust. |

## 4.3 Benchmarking AI Governance in the GCC States

Moving from descriptive comparison to a more integrated scoring, we synthesize how each country fares on the four dimensions that combine MSF insights (Problem/Policy/Politics Streams) with Multi-stakeholder Governance. This benchmarking does not intend to rank countries completely; rather, it illuminates relative strengths and areas for further development:

### 4.3.1 Scoring System

The scoring system assesses each country's performance on the following dimensions derived from the analytical frameworks:

- **Policy Vision and Design**

This dimension assesses clarity, comprehensiveness, and strategic vision of the national AI policies and strategies. It considers how well the policies articulate the problems and opportunities presented by AI and outlines a clear roadmap for AI development and implementation. Derived from (MSF - Problem & Policy Streams).

- **Implementation and Collaboration**

This dimension evaluates the efforts made to implement AI policies and foster collaboration among stakeholders. It considers the presence of dedicated AI units, resource allocation, and initiatives to promote partnerships between government, industry, and academia. Derived from (MSF - Politics Stream & Multi-stakeholder Governance).

- **Stakeholder Engagement**

This dimension assesses the extent and quality of stakeholder participation in AI policy development and implementation. It considers the existence of mechanisms for public consultation, feedback, and involvement in AI-related decision-making processes. Derived from (Multi-stakeholder Governance).

- **Ethical Considerations**

This dimension evaluates the extent to which ethical considerations are integrated into AI governance frameworks. It considers the presence of ethical guidelines, principles for responsible AI development, and measures to address issues such as bias, fairness, and accountability. Derived from (MSF - Multi-stakeholder Governance).

### 4.3.2 Scoring Criteria
- **High:** GCC states with well-defined AI policies, active collaboration among stakeholders, robust stakeholder engagement, and prioritized ethical considerations.
- **Moderate:** GCC states showing progress in the above areas but with some gaps or areas for improvement.
- **Low:** GCC states with early-stage AI governance frameworks lacking clear policy vision, collaborative initiatives, stakeholder engagement, or strong ethical considerations.

### 4.3.3 Benchmarking AI Governance

Figure 6 provides a visual representation of the comparative analysis of AI governance across the GCC states based on the scoring system. Each country is evaluated on the dimensions outlined above, and the scores are plotted on a radar chart to illustrate their relative performance.

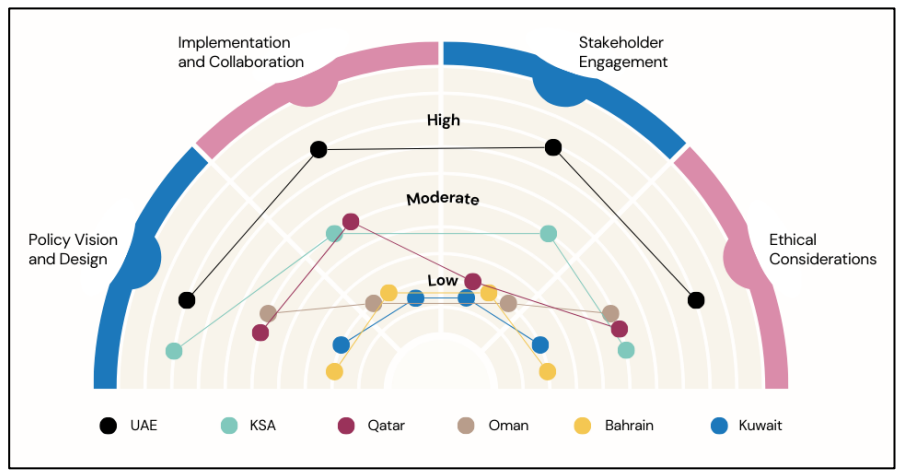

Figure 6: Benchmarking AI Governance in the GCC States

## 5    Discussion

The GCC stands at a crucial point in advancing AI, balancing bold visions with practical governance challenges. Across the region, NASs and ethical guidelines take precedence over formal, binding regulations, reflecting a "soft regulation" paradigm. This approach underscores the GCC's commitment to fostering rapid AI adoption and entrepreneurial innovation, while also highlighting potential risks related to enforcement, alignment with global standards, and the prospect of "ethicswashing" when lofty ethical principles are not paired with robust oversight.

Examining the situation through the lens of the MSF reveals varied degrees of convergence among the key policy streams (problem, policy, and politics). In the UAE and Saudi Arabia, problem definitions, policy solutions, and political commitments align strongly, propelled by top-level leadership and extensive funding. Qatar, Oman, Bahrain, and Kuwait showcase similarly well-articulated problem and policy streams but have yet to fully integrate the political impetus required to convert strategic plans into action. As a result, these countries often face challenges in articulating clear legislative priorities or establishing dedicated AI agencies to drive their respective visions.

At the same time, analysis through a Multi-stakeholder Governance lens underscores the growing involvement of government agencies, large corporations, and academic institutions in shaping GCC AI policies. While bodies like the UAE's Office of Artificial Intelligence, Digital Economy and Remote Work Applications or Saudi Arabia's SDAIA are leading ambitious initiatives, civil society groups and smaller startups continue to have limited roles in determining policy direction. Data protection laws, although in place, also vary in coverage and clarity, leaving gaps that can impede broader stakeholder engagement or undercut public confidence in data handling practices. Strengthening these laws and embedding them within comprehensive AI governance frameworks would foster greater inclusivity and trust among users and innovators alike.

Three challenges consistently stand out across GCC AI strategies: persistent data limitations, the need to cultivate AI-specific human capital, and the task of aligning advanced technologies with the region's cultural and religious values. Although certain initiatives demonstrate promising progress, in particular, large language models tailored to Arabic, specialized AI universities, and references to Islamic ethical concepts in some NAS documents, these remain scattered. High-quality, locally relevant data remain scarce; educational programs are expanding but do not yet meet the market's growing demands; and cultural sensitivities demand nuanced approaches to governance that take account of local norms. Addressing these concerns in an integrated manner is essential to transforming policy aspirations into durable outcomes.

Despite these challenges, the GCC states share clear aspirations related to economic diversification, improved public service delivery, and regional leadership in AI. Many have designated AI development as a cornerstone of broader national visions aimed at reducing reliance on oil and gas. By prioritizing data governance, human oversight, and talent building, they signal a shared resolve to advance AI ethically and responsibly. This common ground opens up possibilities for collaborative R&D initiatives, the creation of cross-border data platforms, and co-development of AI standards relevant to the Arabic-speaking world. Joint efforts can reduce duplication and unify ethical and technical standards.

In comparing GCC governance models with international frameworks, particularly the EU AI Act, a key distinction lies in the stringency of formal regulatory measures. The EU follows a risk-based scheme with prescriptive requirements and clear lines of accountability, while GCC states favor a more flexible and innovation-friendly model. Although this divergence may complicate interoperability or data-sharing agreements, the substantial overlap on ethical principles, particularly transparency, fairness, accountability, and data protection, creates a foundation for future collaboration. Over time, GCC countries may choose to integrate additional risk-management features from the EU and elsewhere, enabling them to preserve local policy preferences while also meeting international standards.

By synthesizing NAS documents, existing legal provisions, and government statements, this study extends AI governance discussions to a region that has, until recently, been underrepresented in academic debates. The Gulf's top-down decision-making style intertwines national visions with the practical rollout of AI in areas like smart government services, finance, and education. Future work should assess long-term outcomes and compare GCC experiences with other emerging markets. Ultimately, ensuring that policy ambitions keep pace with the ethical, cultural, and practical implications of AI will be the central challenge. With continued political commitment, sustained investment in skills and data infrastructure, and more inclusive stakeholder engagement, GCC states stand poised to shape responsible AI innovation on the global stage.

## 6   Conclusion

AI governance in the GCC occurs largely under "soft regulation", prioritizing strategies and ethical guidelines over binding rules. This approach aligns AI agendas with goals for diversification and improved public services, yet the relative absence of formal oversight raises questions about effective enforcement of ethical principles. Although Gulf strategies reference ethics and cultural sensitivity, these remain largely aspirational without robust regulations and stakeholder engagement.

This study addresses gaps in empirical and comparative AI governance research in the GCC. Using a systematic document analysis of NASs, policy statements, and regulatory provisions from 2018 to 2024. Applying the MSF and Multi-stakeholder Governance theory, it illustrates shared aspirations, economic diversification and global leadership, and distinct approaches tied to resource allocation and politics. These findings reveal how high-level visions do or do not translate into workable initiatives, filling a crucial gap in understanding real-world AI policymaking under top-down governance.

Comparative insights show that while the UAE and Saudi Arabia exhibit stronger political will and coherent policy frameworks, other GCC states face challenges in articulating and implementing AI agendas. Key themes include economic diversification, data constraints, talent shortages, and cultural alignment. Civil society and smaller enterprises remain underrepresented, hindering inclusive policies and oversight.

There are notable parallels between GCC strategies and frameworks like the EU AI Act, especially in ethical principles such as transparency, accountability, and fairness. Yet, the Gulf's reliance on non-binding guidelines contrasts with the EU's risk-based regulatory approach; this reliance on non-binding instruments diverges from the EU's risk-based model. Strengthening data protection laws and multi-stakeholder engagement could enhance alignment with global standards for regional AI strategies.

A balanced strategy that encourages AI innovation while safeguarding public trust remains crucial. Future research should examine region-specific ethics, case studies, and policy outcomes over time, longitudinal evaluations of policy impacts, or deeper inquiry into region-specific ethics.

By situating GCC AI governance within a comparative framework, this study helps fill the gap in knowledge regarding how resource-rich, top-down governance structures address responsible AI adoption. In doing so, it underscores the importance of refining governance models for responsible AI adoption and a more inclusive, accountable future.

## Acknowledgments

We wish to acknowledge the support of Kuwait Foundation for the Advancement of Sciences (KFAS).